\documentclass[conference]{IEEEtran}
\IEEEoverridecommandlockouts
\usepackage{cite}
\usepackage{amsmath,amssymb,amsfonts}
\usepackage{algorithmic}
\usepackage{textcomp}
\usepackage{xcolor}
\def\BibTeX{{\rm B\kern-.05em{\sc i\kern-.025em b}\kern-.08em
    T\kern-.1667em\lower.7ex\hbox{E}\kern-.125emX}}
\usepackage[framemethod=TikZ]{mdframed}
\usepackage{fontawesome}
\usepackage{tikz}
\usetikzlibrary{shapes,positioning,shadows,arrows, fit,calc}

\usepackage{lscape}
\usetikzlibrary{mindmap}
\usepackage{tikz}
\usepackage{lscape}
\usepackage{wrapfig}

\begin{document}
\title{Security in Online Freelance Software Development: A case for Distributed Security Responsibility }

% \and
% \and

 \author{\IEEEauthorblockN{ Irum Rauf, Tamara Lopez, Thein Tun,  \\  Marian Petre }
 \IEEEauthorblockA{\textit{The Open University, Milton Keynes, UK} \\
firstname.lastname@open.ac.uk}
 \and
 \IEEEauthorblockN{Bashar Nuseibeh}
 \IEEEauthorblockA{\textit{The Open University, UK} \\
 \textit{Lero, Republic of Ireland}\\
 firstname.lastname@open.ac.uk}}

\maketitle

\begin{abstract}
Secure software is a cornerstone to safe and resilient digital ecosystems. It offers strong foundation to protect users' sensitive data and guard against cyber-threats. The rapidly increasing landscape of  digital economy has encouraged developers from different socio-technical and socio-economic backgrounds to join online freelance marketplaces. While, secure software practices facilitate software developers in developing secure software, there is paucity of research on how freelance developers adhere to security practices and how they can be facilitated to improve their security behavior in under-resourced environments. Moreover, freelance developers are often held responsible for producing insecure code. In this position paper, we review existing literature and argue for the case of distributed security responsibilities in online freelance environment. We propose a research agenda aimed at offering an organized and systematic effort by researchers to address security needs and challenges of online freelance marketplaces. These include: characterising software security and defining separation of responsibilities, building trust in online freelance development communities, leveraging the potential of online freelancing platforms in the promotion of secure software development and building adaptive security interventions for online freelance software development. The research  has the potential to bring forth existing security solutions to wider developer community and  deliver substantial benefits to the broader security ecosystem.

\end{abstract}


\section{Introduction}
Online freelance marketplaces offer advanced systems for remote collaboration, connecting self-employed workers (freelancers) with clients (individuals, small businesses, and large corporations) across the globe \cite{shevchuk2022geography}.  The reported figures by major freelancing platforms suggest that the scale of the global online labor is huge. Being one of the prominent freelance platforms, Upwork reported that more than 145 thousand clients spend over \$ 2.5 billion per year, indicating the platform has significant number of users \cite{Dean_2023}.
% who provide software development services \cite{Dean_2023}. 
Pre-COVID studies estimated that the demand for online freelancing platforms grew by approximately 21 percent from May 2016 to January 2018 with highest demand for software development and technology skills \cite{ kassi2018online}.  COVID-19 has catalysed remote work and the situation looks irreversible with more and more of the workforce adopting remote working model \cite{clarke2023remote}. 

Secure software development is an integral part of software development in today's digitized world with constant security threats looming over businesses and daily lives of individuals. While developer-centered security~\cite{tahaei2019survey} has received much attention in the last decade \cite{smith2018developers}, security in freelance software development has received little attention. 

Below, we highlight the need to investigate the security practices among freelance developers and to motivate the need to provide support to this cohort to develop secure software.

\subsection{Motivation to study Freelance Software Developers for secure software development}
\subsubsection{Existing studies on freelance software developers focus on insecure outcome} 
Existing work on security behavior of freelance developers \cite{naiakshina2019if}, \cite{bau2012vulnerability} and on understanding security in the freelance development ecosystem \cite{ahmed2010agents} notes that freelance software developers produce more insecure code and holds them accountable for it \cite{ahmed2010agents}. However, recent studies (\cite{ita2023lone},\cite{rauf2023security}) attempt to understand why freelance developers produce (more) insecure code. The work of Ryan et al. \cite{ita2023lone} investigates levels of secure coding practices for developers who are under-represented in literature, i.e. isolated developers, open source developers, freelancers and small organisations. They investigate how these cohorts adhere to common security practices. Their empirical findings reveal that these security practices are resource intensive and highlight the  need to target small and under-resourced software development communities with tailored software security advice. The work of Rauf et al. \cite{rauf2023security} suggests that online freelance software development has unique marketplace dynamics that can lead to security compromises. Their work emphasizes the need for tailored security interventions to support freelance software developers working within platforms. 

\subsubsection{Freelance developers can be serious and educated developers} 
The need for offering support to freelance developers to improve their security behavior is exacerbated by the fact that freelance work-model is increasingly being adopted as a serious career - as an alternative to company employment.  The Stack Overflow survey \cite{Stack} reports that nearly 15\% of developers that they surveyed are independent contractors, freelancers, or self-employed, making  online freelance software development(OFSD) a significant part of the software industry. A recent industry report shows that non-temporary freelancers are growing, with  44\% of freelancers saying that they earn more from freelancing than with a traditional job in 2021 \cite{upworkresearch}. 
Moreover, the prevalence of freelancing is increasing among individuals with higher levels of education, while it is declining among those with lower levels of education \cite{upworkresearch}. Similar findings were reported in prior work: an empirical study with freelance developers found that more than 50\% participants had post-graduate education and learnt software development through formal education \cite{rauf2023security}. The study also reported that 90\%  of interviewed freelance developers could be characterized as serious developers who earned regular income from freelancing as full-time or part-time career. These findings about freelance developers from both industry and academia underline the significance of this growing demographic of developers the needs of which should be catered to. 

\subsubsection{Software developed by freelance developers have consequential effects}
Freelance developers are perceived as being non-serious developers who are unreliable \cite{bau2012vulnerability} producing low-quality outputs and showing a lack of commitment around security issues \cite{Hall_2021}. This perception may be grounded on the fact that online freelance marketplaces are open to all kinds of developers - those who know their work well and those who do not. While there are many non-serious developers, online freelancing platforms also host a huge number of serious developers who do a decent job.  This is suggested by the fact that clients increasingly hire from these freelancing platforms and pay them \cite{Dean_2023}. Rauf et al. \cite{rauf2023security} reported that freelance developers do non-trivial jobs, i.e. most of their study participants worked on projects that were customer facing, such a mobile apps, web development, commercial products. 
Moreover, in today's world of digital enhancements, software products increasingly depend on one-another within the software supply chain - - and within which, each job performed forms a significant link.
A clear instance of this is Log4Shell (CVE-2021-44228), a vulnerability found in Log4j, a widely used open-source Java logging tool. This particular flaw was publicly revealed in the latter part of 2021 and was quickly exploited by malicious individuals. By the end of 2022, there were reports indicating that North Korea had utilized this vulnerability to gain initial access to the networks of American energy companies \cite{McCurdy_2022}. 
This indicates that software products developed by freelance developers  have far reaching effects.

\subsubsection{Widespread adoption of easy to use application development frameworks} 
Developing software is no longer the domain of the select few with deep technical skills, training and knowledge. A wide range of people from diverse backgrounds are developing software for smart phones, websites and IoT devices used by millions of people. The rise of easy-to-use development frameworks, such as WordPress have encouraged people from non-technical backgrounds to develop applications that are used by a number of users. To take an example, in an earlier study with freelance software developers\cite{rauf2023security}, participants without a programming background reported that they used WordPress because it offered an easy-to-use interface. However, such frameworks are well-known to attackers for their vulnerabilities \cite{Jackson_2022} - a risk that was perhaps unknown to the clients of freelance developers and of no concern to online freelancing platforms that are only tasked with facilitating transactions.

In this position paper, motivated by the reasons above, we outline a case for identifying roles and responsibilities in online freelance software development and propose a’call-for-action’ to stakeholders of freelancing platforms
to facilitate secure software development practices for this
cohort of developers. We consider it an important step to tackle challenges to writing secure code in online freelance software development platforms that will only magnify with time. Moreover, we see a global presence of developers from different walks of life and different parts of the world. By better leveraging the potential of these freelance developers through tailored security interventions, we can offer developers working in these platforms opportunities to polish their skills and advance their careers by increasing their ability to address vital issues in software engineering in a responsible manner. Moreover, the software development industry can share the benefits of a \textit{skilled workforce} that is globally available on the online freelancing platforms, countering the fast growing need for developers in today's digital economy.

 \section{Distributed Responsibility for Security in Freelance Software Development}

Responsibility in its general sense is often ``concerned with having to answer why one acted as one did'' \cite{lucas1995responsibility}. This often becomes debatable when questioning whether the question is addressed to the right person or not, whether one actually took an action (or not), or whether the question was characterized correctly or not \cite{lucas1995responsibility}. Nonetheless, responsibility is an important concept that helps in holding \textit{someone} accountable for a task that was not performed or not done as it should be. 

The responsibility of security for freelance development is an under-explored area. The work of Ahmed and van den Hoven \cite{ahmed2010agents} consider freelance developers as agents of responsibility in web application development. In the light of existing theories on moral responsibilities of software developers \cite{gotterbarn1995moral} and ethics in information technology \cite{quinn2009ethics}\cite{van1998moral}, their work identifies freelance web developers as ``liable, accountable, blamable, and causally responsible for their work.'' (p.423, \cite{ahmed2010agents}). The work further concludes that `` Freelance web developers are answerable for the possible negative consequences of their actions and omissions.'' (p.423, \cite{ahmed2010agents}). Such viewpoints are exacerbated by empirical studies conducted with freelancers software developers which report that freelance developers lack  responsibility
\cite{ahmed2010agents} and do not attend to security \cite{bau2012vulnerability}.

We find such analysis in line with the sentiment that the \textit{developer is the enemy} \cite{wurster2008developer}. Conversely, aligning with the counterview that the \textit{developer is not the enemy} \cite{green2016developers}, our work shows that freelance developers are not the sole agents of responsibility for secure code. We argue that the responsibility of security in freelance development is better characterized as a \textit{problem of many hands}
\footnote{The term \textit{problem of many hands} is taken from the work of Noorman \cite{sep-computing-responsibility},  wherein, it is discussed as a general issue of determining responsibility in the computing discipline where many parties are involved in the supply chain from developers to end user. In this paper, we discuss the  \textit{problem of many hands} in the context of security in freelance development.}, i.e. it becomes difficult to determine who is responsible for security since multiple entities contribute to the project's security outcome in freelance development making it easy to \textit{assign blame} to someone else for not handling security. This case of assigning blame to other parties is also reported in earlier work \cite{rauf2023security}, wherein some freelance developers consider secure coding responsibility of developer while others consider it responsibility of the client who has to pay for extra effort. 
Below, we outline key stakeholders in online freelance software development and unpick the subtleties of responsibilities of these stakeholders.

% Earlier work \cite{rauf2023security} shows that freelance developers vary in how they attribute responsibility for secure software development in freelance development - some freelance developers consider it a responsibility of freelance developers since they develop the software , other freelancers consider security in software projects a responsibility of the client since client has to pay for the security and give extra time for it. Still others considered it to be the job of auditors or security specialists, if hired by the client, to ensure that software is secure enough, considering security to be a separate feature from development. All of these entities interact with each other on freelancing platforms.  Moreover, the role of freelancing platforms is also an important one as they have the capacity to influence work performance of freelancers \cite{toth2020freelancing}.  Takanen et al. \cite{takanen2004agents} discuss the software vulnerability process with a fictional case study to exemplify the complexity of software vulnerability process and introduce common agents involved in the process. They identify the software vulnerability process through the lens of distribution of responsibility. In this paper, we likewise postulate that security in freelance development is a distributed responsibility.

%\subsection{Context influence Security Behavior of Freelance Developers}

\subsection{Freelance Developer - Responsibilities and Challenges}

The responsibility of freelance software developers in producing secure software is an important one as they use their skills and knowledge to develop applications which have direct or indirect impact on different parts of the society \cite{ahmed2010agents}. They take on the contract and develop software by writing code and/or designing it \footnote{In some scenarios, the developer hired to do the project, may hire other developers to do the task of code development \cite{shevchuk2016heterogeneous}}.
%In freelance environments, freelance developers are often responsible for all the different aspects of the project. Comparing this to developers, in organisational settings developers may work under the influence of product managers or team leads who are often responsible for building business cases for different software features and dealing with the client. 

In order to hold someone accountable for a job, it is important that the one being questioned has control over his/ her action \cite{lucas1995responsibility}. Research \cite{rauf2023security} suggests that freelance developers are not oblivious to their responsibility and try to find a \textit{work around} where they are challenged. However, freelance developers are often constrained in their jobs by different socio-technical factors, such as multivalent nature of security, relationships with client, algorithms of freelancing platforms, and choice of different development frameworks.  Below we discuss these briefly. 

% Below, we discuss some of the responsibilities that emerged from our work with freelance software developers \cite{rauf2023security}, the challenges they face in fulfilling these responsibilities and what are the works around that freelance developers take to tackle these challenges. 
%- however these works ar that responsibilties how developers perceive this responsibility, what are some of the basic responsibilities of freelance developers, and what are the challenges they face in fulfilling their responsibilities.

%Most of the freelance developers consider security responsibility of the developer and . However, they vary in how they define it and and have an unclear understanding of their responsibility.

In order to hold someone accountable for a job, it is important to ask the right question, i.e. is the question characterised correctly or not? \cite{lucas1995responsibility}. Freelance software developers are held responsible for doing security \cite{ahmed2010agents}. However, security vulnerabilities can be of varying nature. It  ``can be a lacking security requirement (e.g. lack of, or improper authentication, encryption, ...), or a development error in the software (e.g. buffer overflow, race condition, ...).'' (p.93, \cite{ takanen2004agents}). Some security requirements are well know and hence many developers consider them as basic security, e.g. authentication, password hashing and encryption of sensitive information \cite{rauf2023security}.Due to multivalent nature of security \cite{rauf2022challenges} and diverse skill set of freelance developers \cite{ita2023lone}, participants have different perceptions of security \cite{rauf2022chase}. Different perceptions on basic security result in false perceptions in developers that they are handling (or not handling) security~\cite{naiakshina2019if}. Software security researchers need to explicitly define tangible characteristics of security that developers should adhere to. 

Some freelancers opt for popular development frameworks because they are easy to use not requiring expert programming skills \cite{rauf2023security}. Some of these frameworks maybe insecure \cite{Jackson_2022} but offer (paid) secure plugins. However, clients are not always willing to pay \cite{rauf2023security}. Moreover, some freelancers find it hard to stay updated with various security plugins of such frameworks \cite{rauf2023security}. Here again we notice that freelancers are aware of shortcomings of the frameworks they use with some switching to another development framework and others tend to hide URLS in an effort to avoid attention of the attackers \cite{rauf2023security}. Other freelancers, who heavily rely on development frameworks tend to stay updated with their frameworks as they do not have time to stay updated with changes in security landscape in general \cite{rauf2023security}. 
% Developers feel comfortable using development frameworks since they take care of many low-level details and have security built-in into their development infrastructure. The work of Rauf et al. \cite{rauf2022chase} highlights that  while security may not be a \textit{tangible} quality to developers, when it is packaged with other code quality attributes that are visible to developers in their immediate contexts, it is adopted. We recommend, that freelance developers should ask the client to use sec
% ure frameworks, if they suggest otherwise. Furthermore, developers should keep themselves updated with new versions of development framework which often update their frameworks to reflect state of the art in security and other technology developments.

% The use of mature and popular development frameworks and libraries is encouraged by security experts compared to writing own solutions \footnote{\url{https://www.securecodewarrior.com/}} . 

Moreover, empirical studies \cite{rauf2023security} suggest that freelance developers consider it responsibility of freelancers to initiate discussion on security with the client and inform  about any security issues to non-technical clients in particular. However, developers find it difficult to discuss security issues with non-technical clients who think freelancers are finding ways to make extra money. Henceforth, some freelancers try to work with only technical clients who understand the technicalities of software projects, or they work around by developing long- term working relationship with their client to infuse trust in their relationships. Nonetheless, FL developers who are new to online platforms, struggle to select right clients. Algorithms in online platform provide greater visibility to developers who have done more projects and have good rating from clients. Thus, these freelancers may have to compromise on security in order to complete a reasonable number of projects with clients who don't take security seriously.Only when they have a stronger profile, they are in a better position to select clients who understand technical requirements of the project and give extra time and money for secure development. A recent study by Munoz et al. \cite{munoz2022platform} offer similar insights on ``how online freelancer’s identity presentation is constrained by the structuring of their profile, the ratings and client feedback, the algorithms used by the digital platform, and platform’s terms of use'' (p.1).  The study reports that freelance workers realize how these platforms control their identity are resist their deconstructed identity by the online platforms.

% \subsubsection{Characterising basic security}

% Earlire studies on freelance software developers [\cite{rauf2022chase} and \cite{rauf2023security}) find many instances in which freelance developers address such security requirements as basic security that a developer should address invariably. On the other hand, we also found instances of freelance developers, reported in aforementioned studies,  where freelance developers only consider development errors as basic security issues that a developer is answerable for. Thus, there is a lack of clear understanding on what a developers' responsibility is in terms of basic security that a developer should address. 

Lack of adoption of common security practices in this cohort of developers is also a challenge \cite{ita2023lone}.  Earlier study with freelance web application developers \cite{rauf2022chase} showed that many freelancers are unaware of OWASP top 10 list of web application vulnerabilities \cite{OWASPSec30:online} and more recent study \cite{ita2023lone} showed that the use of automated security tools is very low in freelance developers which can be of most benefit to \textit{under-resourced} developers.
\subsection{Client - Responsibilities and Challenges}
 
Clients are an important stakeholder of freelance development as they hire a freelance developer and pay for the project. In this section, we outline the responsibilities of the clients to encourage them to take a responsible role in freelance development. In the presence of explicit security requirements, (freelance) developers \textit{tend to} produce secure software as they are primed to think of security \cite{naiakshina2019if} and also the software product can be validated against security requirements \cite{haley2008security}. However, clients may not always have a technical background and security may not be on the top of their head. In such scenarios, clients find it difficult to trust freelance developers who ask for extra money for secure development \cite{rauf2023security}.  Studies report that \textit{trust} as an important factor in the client-freelancer relationship influences how security is handled in freelance software projects \cite{rauf2023security}.

Moreover, clients are often advised to hire good developers if they want a secure product, which is often translated to hiring expensive developers but is not always the case \cite{naiakshina2019if}\cite{bau2012vulnerability}. It is important to help clients, who are interested in developing good quality secure software to find the right talent in freelance online software development market. While the clients are willing to pay extra, Rauf et al. \cite{rauf2023security} report that perceptions on payment for security vary. While some freelance developers charge extra for secure development, others may not and still do secure coding considering it part of development.  Furthermore, the current feedback mechanism in freelancing platforms rank freelancers on positive feedback from clients mainly on meeting project timeline and good communication. This makes it challenging for clients, especially the non-technical client, to hire the \textit{right} developer. 

Moreover, technology naive clients find it difficult to have meaningful conversations with the freelance developers which results in compromise on security. Some freelancers avoid security because clients never ask for it \cite{rauf2023security}. Clients should explicitly discuss developers' security perceptions to ensure security is address in their projects. Additionally, responsibilities in teams are often not explicitly defined in remote teams \cite{rauf2023security}. This exacerbates the \textit{problem of many hands} with freelance developers often holding someone else responsible for security in the project. These challenges require that clients should be facilitated with security interventions to raise their awareness of insecure software and understand the business case of security in software. Additionally, platforms should provide easy to understand security information to clients to have a meaningful conversations with freelance developers. Clients should also make it explicit to freelance developers working in online teams if there is \textit{someone else} responsible for security.

\subsection{Freelancing Platforms -  Responsibilities and Challenges}
The role of freelancing platforms is  an important one as they have the capacity to influence work performance of freelancers \cite{toth2020freelancing}.
  Although moral responsibilities have in general revolved around the role of humans, with the prevalence of technology, the human activities cannot be fully understood without a reference to technological artifacts \cite{waelbers2009technological}. The online freelancing platforms which are actively used by developers around the world are \textit{sociotechnical systems}. Based on the work of Bijker et al. \cite{Bijker1987social},  sociotechnical systems  are defined by Noorman  \cite{sep-computing-responsibility} as the systems in which ``tasks are distributed among human and technological components, which mutually affect each other in contingent ways'' (p.1.). Freelancing platforms act as ``active mediators'' \cite{kroes2014moral} and have the potential to promote security in freelance software development. Verbeek \cite{verbeek2006materializing} highlights technological artifacts as ``active mediators'' that ``actively co-shape people's being in the world: their perception and actions, experience and existence''( p. 364). In recent years, we have seen skyrocketed rise in the business value of freelancing platforms  \footnote{https://www.zdnet.com/finance/upwork-delivers-uneven-q3-but-touts-904-million-of-total-value-sourced-from-platform/} with  a sharp increase in freelance workforce (UK alone has seen an increase of 46\% from 2008 to 2017 \cite{jenkins2017exploring} ). We postulate that freelancing platforms hold a pivotal position to influence behavior of clients and developers by offering (security) interventions and fulfill their social responsibility as active mediators. This is in line to the work of Gottenbarn \cite{gotterbarn2001informatics} -  according to Gottenbarn considering technological artifacts as ethically neutral is a misplaced belief and there can be detrimental consequences of missing the broader context in which the technologies sit in. Unfortunately, despite the pivotal position that freelancing platforms hold in freelance software development ecosystem, to the best of our knowledge we did not find any research in how freelancing platforms can facilitate and promote security culture in freelance software development environment. While developer centered security is an active research areas \cite{rauf2021adaptive} with researchers and practitioners studying and facilitating security culture in software companies \cite{tuladhar2021analysis} and open-source communities \cite{wen2019empirical} and also investigating security responsibilities in software companies \cite{charles2022responsibility}, there is a need to focus research efforts on understanding the nuances of security responsibility in online freelance software development and the role of freelancing platforms in promoting security responsible behavior.

\section{A Research Agenda for Promoting Secure Software Development in Online Freelance Environment}
Our analysis of existing literature suggests the need for a holistic look at secure coding behavior of freelancers and understanding the complex the socio-technical context they work in. Recent studies identify the unique marketplace dynamics of freelance software developers and the the nuances of security perceptions held by them \cite{rauf2023security}. Furthermore, research identifies that common security practices for secure software developed are insufficient for under-resourced developers and highlights the need for tailored security interventions for them\cite{ita2023lone}. 

Going forward we outline our research agenda and organize our suggestions into four areas to investigate: 

\subsection{Characterising software security and defining separation of responsibilities}

In order to encourage consistent understanding of secure software development and facilitate separation of responsibilities, we postulate characterising security to identify basic and advanced security with separation of responsibilities, We suggest conducting empirical studies with professional developers, security experts and freelance developers to understand what they think is basic security that should be done as part of development without explicit security requirements. The thematic analysis on how freelance developers define basic and advance security \cite{rauf2023security} can be a good starting point.  We then suggest use of authoritative sources to characterise security and provide a draft of basic responsibilities of a developer.

\begin{mdframed}[roundcorner=10pt]
\textit{Key Research Questions:} 
\begin{itemize}
   \item How do developers and security specialists define  \textit{basic} security responsibilities?
\item What do security experts consider part of secure software development?
\item How do security responsibilities vary with programming languages and development frameworks?
\item How can we provide \textit{separation of security responsibilities} and get consensus on it? 
\end{itemize}

\end{mdframed}

\subsection{Building trust in online freelance development communities}
% How can trust relationship be build between clients and freelancers? 
Clients and freelance developers work together to produce a secure software. However, mistrust between the two can result in security compromise. We encourage multidisciplinary research to investigate theories of trust from behavioral sciences and use them to build trust in software communities for security.  Toth et al. \cite{toth2020freelancing} conducted a survey with 127 freelancers to explore the relationship between virtual community trust, work engagement. \textit{Work-engagement} has a strong link to meaningful work \cite{macey2008meaning} and is defined as : ``“a positive, fulfilling work-related state of mind that is characterized by vigor,
dedication and absorption” (p. 74, \cite{schaufeli2002measurement}) and person–job fit  is described as  a match between personal abilities and demands of the job \cite{toth2020freelancing}. The works suggests that trust in digital communities positively affects both the work-engagement and person-job fit.  "Freelancing platform can improve work performance through person–job fit by assisting in the creation of trust among members of their platforms'' (p.1, \cite{toth2020freelancing}). Recent study by Bianca et al. \cite{trinkenreich2023belong} investigates factors that influence sense of belong of developers to a virtual community. The sense of belonging to a community retain contributors and improve project sustainability.  It is important to examine the factors that impact the sense of virtual community among freelance developers in online freelance marketplaces. Furthermore, these platforms should incorporate these factors to enhance project sustainability and ensure the retention of freelance developers.

Furthermore,  we advocate the use of rich resources developed by academia and industry on computing code of ethics, and security community to offer induction courses to freelance developers when onboarding online freelance platforms. While, these courses may not be mandatory but developers who attempt them should get rewarded via badges or higher rank in search algorithms. Rogerson suggests that ``Codes of ethics and practice can be enormously powerful if used proactively.'' \cite{rogerson2002software}.  Software Engineering Code of Ethics and Professional Practices  \cite{gotterbarn2014software}  can be used as springboards to offer membership/ licensing by institutions \cite{ahmed2010agents}. Freelance developers should be made aware and encouraged to take membership of professional bodies that promote responsible behavior. They should ``strive to become members of international associations or community of computing'' (p.422, ,\cite{ahmed2010agents}) and display it on their profile to stand-out from the crowd. Moreover, the freelancing platforms should adjust their algorithms to highlight the profiles of developers who advocate responsible behavior and display such badges.

\begin{mdframed}[roundcorner=10pt]
\textit{Key Research Questions:} 
\begin{itemize}
   \item How can we use theories of building trust in communities from behavioral sciences in online freelance communities?
   \item How can we leverage the extensive body of work on ethics in software engineering and utilise it to encourage responsible behaviour among developers?
    \item How can freelancing platforms be encouraged to update their algorithms to highlight security responsible behavior?
\end{itemize}

\end{mdframed}

\subsection{Leveraging the potential of online freelancing platforms in the promotion of secure software development}
% How can we leverage on the potential of freelancing platforms in secure software development? 
Freelancing platforms have the potential to influence freelancers behavior and security culture in freelancing communities in a number of ways. However, the challenge is how to onboard freelancing platforms on this and build a compelling business case for security to them. Moreover, onboarding freelancing platforms on building security culture in freelancing platforms also comes as a moral and social responsibility that they are accountable for. 

 Online freelance marketplace hold a unique and pivotal position in today's digital landscape to educate and influence developers who are under-resourced and come from deprived economy.   Moreover, by offering security interventions in proximity to developers via online freelance marketplaces, it is possible to enhance the skills and conduct of this group of developers. This approach can effectively address the growing demand for responsible developers in the present digital economy.

Research identifies that online freelance platforms influence and control identities of freelancers \cite{munoz2022platform}. There is an immediate need to highlight the power that online freelance marketplaces hold in digital economy and over the careers of freelance developers. However, \textit{with great power comes responsibility}. These responsibilities should come forth and researchers and security industries need to construct a convincing proposal for security to these platforms to get them onboard on promoting secure and responsible behavior among freelance developers. 

Our empirical work \cite{rauf2023security} also suggest many freelance developers also work in companies and opt for freelancing as a part-time job, or switch often between the two. Existing studies also suggest that workers may combine employment statuses by having multiple jobs \cite{shevchuk2016heterogeneous}. The empirical study  with freelancers , reported by Shevchuk and Strebkov \cite{shevchuk2016heterogeneous}, report how individuals' work values differ in their self-employment situations. The different value sets is also evident in developers wherein Rauf et al. \cite{icsew2020} reported that a developer shifted on his value-set depending on whether he is working on a project for the company or for himself. These differences in developers’ value-sets as they switch their working hat is noteworthy in the realm of security in freelance community. We consider it crucial to investigate the impact of mentorship for security through freelancing platforms on the security mindset of developers who work with companies lacking a security culture. This research direction holds significant importance.

Freelance platforms also have great potential to facilitate researchers in conducting empirical studies. Danilova et al. \cite{danilova2020replication} suggest that use of online freelancing platforms provides ecological validity for online security developer studies. However, experience of researchers (e.g., \cite{rauf2022challenges}, \cite{danilova2020replication} and \cite{rauf2023security)} with freelance platforms suggest that freelance platforms do not encourage researchers to recruit freelancers directly for research studies. Rauf et al. \cite{rauf2022challenges} report recruiting relatively large number of freelance software developers for a research study through a non-friendly user interface requiring them to create separate jobs and contracts for each individual freelance developer (reported in \cite{rauf2022challenges}. While it made the job very lengthy and exhausting (considering hiring of at least 124 freelance developers \cite{rauf2022chase}),the requirement of job also required that freelancing platforms deduct a considerable amount from the paid amount for each study participant which may not scale well for limited research funding. The members of freelance platforms are not allowed to take payment outside the platform as the correspondence between the client and a freelancers are often checked and then penalised if there is a conversation on payment through other means \cite{rauf2022challenges}.  Moreover, the study \cite{rauf2022challenges} report rejection from some freelancing platforms who did not approve research study considering it unsuitable for their platform. 

%Additionally, the search interfaces are not intuitive for researchers who often need to recruit a large number of participants for the same task. 
 \begin{mdframed}[roundcorner=10pt]
\textit{Key Research Questions:} 
\begin{itemize}

\item How can we effectively advocate for freelancing platforms to promote security interventions in a compelling business case?
\item How can we highlight the case of moral/ social responsibility of freelancing platforms? 
\item  How does \textit{value transfer} occur  between the freelance development and company work?
\item Can mentorship in freelance environment propagate to company practices ? 
\item Which recruiting strategy is effective in recruiting freelance software developers for research studies?
\item How can we create a business case for freelance platforms to promote and support research on freelance software development?

\end{itemize}

\end{mdframed}

\subsection{Building adaptive security interventions for online freelance software development}
 % How to build adaptive security interventions to support distributed security responsibility in freelance development environment?
``Adaptive security interventions take the socio-technical context into account, and therefore respond to the different security needs of the developer '' (p.25, \cite{rauf2021adaptive}). We postulate that \textit{distributed security responsibility}, wherein all the involved parties are aware of their responsibilities and comply with them, can be best done with adaptive security interventions. These adaptive security interventions should facilitate freelancers and clients in developing secure software. Clients have their own set of requirements such as awareness of negative consequences of software vulnerabilities and guidelines on how to recruit \textit{right} developer in a given domain and development framework. Similarly, freelancers also work under varying needs and socio-technical settings. Their intervention needs can range from gaining familiarity of different types of security interventions \cite{rauf2022chase} depending on domain, programming language,  development frameworks, and developers' socio technical environment such as working alone or in team, and support in making a business case for security for different types of clients. These interventions need to be designed in an cost and time effective manner to capture attention of freelancers who are often time poor. We also believe that security interventions should focus on positive responsibility \cite{gotterbarn2001informatics}, i.e. focusing on what ought to be done and provide incentives to freelance developers rather than on blaming or punishing them for irresponsible behavior. 

% \noindent \textit{Recommendation:} Research efforts should focus on identifying factors that can help clients in identifying freelance developers who produce secure software. For example,  security researchers can provide an unambiguous separation of security responsibilities between freelance developers and the clients. This should be advertised  closer to their environment, i.e., propagating them through freelancing platforms. Freelancing platforms can promote profiles of freelance developers  who are members of  the ethics committees. Induction courses on responsibilities of a developer can be designed when joining freelancing platforms with positive incentives such as badges that filter them when clients search for responsible software developers.
\begin{mdframed}[roundcorner=10pt]
\textit{Key Research Questions:} 
\begin{itemize}
\item How can we design security interventions to encourage security responsible behavior as positive responsibility in freelance development?
    \item How can we design interventions that can help freelance developers make security a selling point for clients?
\item How can we design adaptive security interventions for different types of developers and clients in a cost and time effective manner?
\end{itemize}

\end{mdframed}

\section{Conclusion}

In this position paper, we advocate the need for organized and systematic effort by researchers to address security needs and challenges of online freelance marketplaces. Based on understanding of existing literature, rapid adoption of freelance work model and exponential growth in the revenue of online freelance marketplaces, we highlight the case of distributed security responsibility among different stakeholders of online freelance software development. The unique dynamics of online freelance marketplaces offers interesting challenges to advancing research in this domain, but it has the potential to bring forth existing security solutions to wider developer community and  deliver substantial benefits to the broader security ecosystem.

\bibliographystyle{IEEEtran}
\bibliography{fl22}

%  \appendix{Appendix}
% \section{Appendix}

% \begin{table*}[]
% \begin{tabular}{|l|l|l|l|l|l|l|l|l|l|}
% \hline
% \multicolumn{2}{|c|}{Gender} & \multicolumn{2}{|c|}{Education}  &  \multicolumn{3}{c|}{Age (years)}  & \multicolumn{3}{c|}{Location}   \\ \hline
% Male & Female & Masters & Bachelors & 18 - 24 & 25 - 34 & 35 - 44 & India & Pakistan & Others\\ \hline
%  95\% & 5\%  & 55 \% & 45 \% & 10 \% &85 \% & 5 \% & 55 \% & 30 \% & 15 \% \\ \hline

% \end{tabular}
% \caption{Participants' Demographics by Gender, education and geographical Distribution }
% \end{table*}

% \begin{table*}[]
% \begin{tabular}{|l|l|l|l|l|l|l|l|l|l|}
% \hline
%  \multicolumn{2}{|c|}{Prog. Exp. (years)}  &  \multicolumn{2}{c|}{Learning Background} &  \multicolumn{2}{c|}{FL Exp. (years)} &  \multicolumn{2}{c|}{FL projects*}& \multicolumn{2}{c|}{Income}  \\ \hline
% 2 to 5 &  35 \% & As part of education  & 60\%  & $ \leq $ 2 & 5 \% & 2-5 & 70\% & Part-time  (regular) & 50\% \\ \hline
% 6- 10 & 50\% & Self-taught & 30\% & 2-5 & 75 \% & 6 - 10  & 15\% & Full-time (Primary income)  & 40\% \\ \hline
%  10+ & 15\%  & Training at work & 10 \% & 6-10 & 15 \% & >11 & 15\% & Occasionally & 10\%\\ \hline

% \end{tabular}
% \caption{Participants' Professional Demographics (Here: Prog. Exp. = Programming Experience, FL = Freelancer, Exp. = Experience, * in last year)}
% \end{table*}

\end{document}